\documentclass[a4paper,11pt]{article}

\usepackage{contribution}
\usepackage{epsfig}
\usepackage{graphicx}

\newcommand{\weblink}[2][]{%
    \ifthenelse{\equal{#1}{}}%
    {\textnormal{\url{#2}}}%
    {\textnormal{\href{#2}{#1}}}%
}

\newcommand{\acknowledgements}[1]{%
  \bigskip\bigskip
  \textsf{\textbf{\Large Acknowledgements}} \\[2ex]
  {#1}
  \bigskip
}

\def\beq{\begin{equation}}
\def\eeq#1{\label{#1}\end{equation}}
\def\eeqn{\end{equation}}

\def\beqa{\begin{eqnarray}}
\def\eeqa#1{\label{#1}\end{eqnarray}}
\def\eeqan{\end{eqnarray}}

\let\bar=\overbar

\def\Dslash{\not{\hbox{\kern-4pt $D$}}}
\def\dslash{\not{\hbox{\kern-2pt $\del$}}}

\def\msb{{\bar{\ssstyle M \kern -1pt S}}}

\newcommand{\contribution}[7][]{%
  \clearpage
  \thispagestyle{plain}
  \ifthenelse{\equal{#1}{}}
  {\hypersetup{pdftitle={#2}}}
  {\hypersetup{pdftitle={#1}}}
  \hypersetup{pdfauthor={{#3} {#4}}}
  {\centering\normalfont\LARGE\bfseries\sffamily #2 \par\nobreak}
  \lhead{}
  \chead{%
    \textit{\footnotesize XIV International Conference on Hadron Spectroscopy
      (\weblink[\textit{hadron2011}]{http://www.hadron2011.de}), 13-17 June 2011, Munich, Germany}%
  }
  \rhead{}
  \bigskip
  \begin{center}
    {#3} {#4}\ifthenelse{\equal{#6}{}}{}{\footnote{\weblink[#6]{mailto:#6}}}
    \ifthenelse{\equal{#7}{}}{}{#7} \\
    \textit{#5}
  \end{center}
  \bigskip
}

\renewcommand{\abstract}[1]{%
  \begin{center}
    \begin{minipage}{0.85\textwidth}
      \begin{footnotesize}
        #1
      \end{footnotesize}
    \end{minipage}
  \end{center}
  \bigskip
}

\makeatletter
\@ifundefined{c@affiliation}%
{\newcounter{affiliation}}{}%
\makeatother
\newcommand{\affiliation}[2][]{\setcounter{affiliation}{#2}%
  \ensuremath{{^{\alph{affiliation}}}\text{#1}}}
\def\tstrutb{\vrule height2.5ex depth0pt width0pt} % used in tables

\begin{document}

% % % % % % % % % % % % % % % % % % % % % % % % % % % % % % % % % % % % % % % % 
% your proceedings
%%%%%%%%%%%%%%%%%%%%%%%%%%%%%%%%%%%%%%%%%%%%%%%%%%%%%%%%%%%%%%%%%%%%%%%%%%%%%%%%
%
% template for hadron2011 contribution
%
% please do not rename this file
%
% to create document run
%
%     pdflatex hadron2011.tex
%
%%%%%%%%%%%%%%%%%%%%%%%%%%%%%%%%%%%%%%%%%%%%%%%%%%%%%%%%%%%%%%%%%%%%%%%%%%%%%%%%
{  % do not remove

%%%%%%%%%%%%%%%%%%%%%%%%%%%%%%%%%%%%%%%%%%%%%%%%%%%%%%%%%%%%%%%%%%%%%%%%%%%%%%%%
% please define your macros here

%
%%%%%%%%%%%%%%%%%%%%%%%%%%%%%%%%%%%%%%%%%%%%%%%%%%%%%%%%%%%%%%%%%%%%%%%%%%%%%%%%

%%%%%%%%%%%%%%%%%%%%%%%%%%%%%%%%%%%%%%%%%%%%%%%%%%%%%%%%%%%%%%%%%%%%%%%%%%%%%%%%
% define title, author, and address
\contribution[Weak $B$ Decays into Orbitally Excited Charmed Mesons]
{Weak $B$ Decays into Orbitally Excited Charmed Mesons}
{J.}{Segovia}  % presenter of the talk/poster
{\affiliation[Departamento de F\'isica Fundamental and IUFFyM]{1} \\
 Universidad de Salamanca, E-37008 Salamanca, Spain}
{segonza@usal.es}
{\!\!$^,\affiliation{1}$,
C. Albertus\affiliation{1}, 
D.R. Entem\affiliation{1},
F. Fern\'andez\affiliation{1},
E. Hern\'andez\affiliation{1},
and M.A. P\'erez-Garc\'ia\affiliation{1}}
%
%%%%%%%%%%%%%%%%%%%%%%%%%%%%%%%%%%%%%%%%%%%%%%%%%%%%%%%%%%%%%%%%%%%%%%%%%%%%%%%%

%%%%%%%%%%%%%%%%%%%%%%%%%%%%%%%%%%%%%%%%%%%%%%%%%%%%%%%%%%%%%%%%%%%%%%%%%%%%%%%%
% abstract
\abstract{%
The BaBar Collaboration has recently reported branching fractions for
semileptonic decays of the $B$ meson into final states with charged and neutral
$D_{1}(2420)$ and $D_{2}^{\ast}(2460)$, two narrow orbitally excited charmed
mesons. We evaluate these branching fractions within the framework of a
constituent quark model in two steps, one which involves a semileptonic decay
and the other one mediated by a strong process. Our results are in agreement
with the experimental data.}
%
%%%%%%%%%%%%%%%%%%%%%%%%%%%%%%%%%%%%%%%%%%%%%%%%%%%%%%%%%%%%%%%%%%%%%%%%%%%%%%%%%

%%%%%%%%%%%%%%%%%%%%%%%%%%%%%%%%%%%%%%%%%%%%%%%%%%%%%%%%%%%%%%%%%%%%%%%%%%%%%%%%%
% main text
% for short contributions sections are optional
\section{Introduction}

Different collaborations have recently reported semileptonic $B$ decays into
orbitally excited charmed mesons providing detailed results of branching
fractions~\cite{belle,aubert09}. These data offer new theoretical possibilities
to test meson models as far as they include a weak decay followed by a strong
one.

All these magnitudes can be consistently calculated in the framework of
constituent quark models because they can simultaneously account for the
hadronic part of the weak process and the strong meson decays. In this context
meson strong decay has been described successfully in phenomenological models,
like the $^{3}P_{0}$ model~\cite{mic69_1} or in microscopic models (see
Refs.~\cite{Eichten78,Swanson96}). The matrix element for the weak process
factorizes into a leptonic and a hadronic part. It is the hadronic part that
contains the non-perturbative strong interaction effects and we shall evaluate
it within the constituent quark model (CQM) of Ref.~\cite{Vijande2005} which
successfully describes hadron phenomenology and reactions. Details of the
calculation can be found in Ref.~\cite{Segovia2011}.

\section{Theoretical framework}

\subsection{Constituent quark model}

Spontaneous chiral symmetry breaking of the QCD Lagrangian together with the
perturbative one-gluon exchange (OGE) and the non-perturbative confining
interaction are the main pieces of potential models. Using this idea, Vijande
{\it et al.}~\cite{Vijande2005} developed a model of the quark-quark interaction
which is able to describe meson phenomenology from the light to the heavy quark
sector. Further details can be found in Ref.~\cite{Vijande2005}. 

In order to find the quark-antiquark bound states, we solve the Schr\"odinger
equation by Rayleigh-Ritz variational principle. We use the Gaussian Expansion
Method~\cite{Hiyama2003} that provides enough accuracy and makes the subsequent
evaluation of the decay amplitude matrix elements easier.

Model parameters are given in Ref~\cite{Segovia2008}.

\subsection{Weak and strong decays}

In the weak decay we have a $\bar{b} \to \bar{c}$ transition at the quark level
and we need to evaluate the hadronic matrix elements of the weak current
\begin{equation}
J^{bc}_\mu (0)=\bar{\psi}_b(0)\gamma_\mu(I-\gamma_5)\psi_c(0).
\label{eq:current1}
\end{equation}

The hadronic matrix elements involved in these processes can be parametrized in
terms of form factors. The expression of the hadron tensor in the helicity
formalism~\cite{ivanov} has been calculated following Ref.~\cite{hnvv06}.

To describe the meson decay process $A\rightarrow B+C$, the $^{3}P_{0}$ decay
model assumes that a quark and an antiquark are created with vacuum quantum
numbers. The created $q\bar q$ pair together with the $q\bar q$ pair from the
initial meson regroups in the two outgoing mesons via a quark rearrangement
process. For the $^{3}P_{0}$ decay model, the interaction Hamiltonian is given
by
\begin{equation}
H_{I}=g\int d^{3}x \bar{\psi}(\vec{x})\psi(\vec{x})
\end{equation}
where $g$ is related to the dimensionless constant giving the strength of the 
$q\bar q$ pair creation from the vacuum as $\gamma=\frac{g}{2m_{q}}$.

In the microscopic decay models, the strong decays are driven by the
interquark Hamiltonian which determines the spectrum. In our case we have the
one-gluon exchange and a mixture of scalar and vector Lorentz confining
interactions appearing as the kernels. These interactions and their associated
decay amplitudes are undoubtedly all present and should be added coherently. The
Hamiltonian of the interaction can be written as
\begin{equation}
H_{I}=\frac{1}{2}\int 
d^{3}\!xd^{3}\!y\,J^{a}(\vec{x})K(|\vec{x}-\vec{y}|)J^{a}(\vec{y}),
\label{Hint}
\end{equation}
where current $J^{a}(\vec{x})$ in Eq.~(\ref{Hint}) is assumed to be a
color octet. Calculation details referred to the microscopic model can be found
in Ref.~\cite{SegoviaHadron}.

\section{Results}

The final results and their comparison with the experimental data are given in
Table~\ref{tab:experiment2}. Both $^{3}P_{0}$ and microscopic models predict
similar branching ratios. The predictions for the $B\to D_{1}l\nu_{l}$ and
$B\to D_{2}^{\ast}l\nu_{l}$ are in good agreement with the latest experimental
measurements by the BaBar Collaboration. They are significantly smaller than the
Belle data, though.

\begin{table}[t!]
\begin{center}
\begin{tabular}{lcccc}
\hline
\tstrutb
& Belle~\cite{belle} & BaBar~\cite{aubert09} & $^{3}P_{0}$ & Mic. \\
& $(\times10^{-3})$ & $(\times10^{-3})$ & $(\times10^{-3})$
& $(\times10^{-3})$ \\
\hline
\tstrutb
$D_{1}(2420)$ & & & & \\
${\cal B}(B^+ \to \bar{D}^0_1 l^+ \nu_l){\cal B}(\bar{D}^0_1\to D^{\ast-}\pi^+)$
& $4.2\pm0.7\pm0.7$ & $2.97\pm0.17\pm0.17$ & $2.57$ & $2.57$ \\
${\cal B}(B^0 \to D^-_1 l^+ \nu_l){\cal B}(D^-_1\to \bar{D}^{\ast0}\pi^-)$ &
$5.4\pm1.9\pm0.9$ & $2.78\pm0.24\pm0.25$ & $2.39$ & $2.39$
\\
\hline
\tstrutb
$D_{2}^{\ast}(2460)$ & & & & \\
${\cal B}(B^+ \to \bar{D}^{\ast0}_2 l^+ \nu_l){\cal B}(\bar{D}^{\ast0}_2\to
D^{-}\pi^+)$ & $2.2\pm0.3\pm0.4$ & $1.4\pm0.2\pm0.2^{(\ast)}$ & $1.43$ &
$1.47$ \\
${\cal B}(B^+ \to \bar{D}^{\ast0}_2 l^+ \nu_l){\cal B}(\bar{D}^{\ast0}_2\to
D^{\ast-}\pi^+)$ & $1.8\pm0.6\pm0.3$ & $0.9\pm0.2\pm0.2^{(\ast)}$ & $0.79$ &
$0.75$ \\
${\cal B}(B^+ \to \bar{D}^{\ast0}_2 l^+ \nu_l){\cal B}(\bar{D}^{\ast0}_2\to
D^{(\ast)-}\pi^+)$ & $4.0\pm0.7\pm0.5$ & $2.3\pm0.2\pm0.2$ & $2.22$ &
$2.22$ \\
${\cal B}(B^0 \to D^{\ast-}_2 l^+ \nu_l){\cal B}(D^{\ast-}_2\to
\bar{D}^{0}\pi^-)$ & $2.2\pm0.4\pm0.4$ & $1.1\pm0.2\pm0.1^{(\ast)}$ & $1.34$
& $1.38$ \\
${\cal B}(B^0 \to D^{\ast-}_2 l^+ \nu_l){\cal B}(D^{\ast-}_2\to
\bar{D}^{\ast0}\pi^-)$ & $<3$ & $0.7\pm0.2\pm0.1^{(\ast)}$ & $0.74$ & $0.70$
\\
${\cal B}(B^0 \to D^{\ast-}_2 l^+ \nu_l){\cal B}(D^{\ast-}_2\to
\bar{D}^{(\ast)0}\pi^-)$ & $<5.2$ & $1.8\pm0.3\pm0.1$ & $2.08$ &
$2.08$ \\
${\cal B}_{D/D^{(\ast)}}$ & $0.55\pm0.03$ & $0.62\pm0.03\pm0.02$ & $0.65$ &
$0.66$ \\
\hline
\end{tabular}
\caption{\label{tab:experiment2} Most recent experimental measurements reported
by Belle and BaBar Collabo\-rations and their comparison with our results. The
symbol $(\ast)$ indicates the estimated results from the original data using
$B_{D/D^{(\ast)}}$.}
\end{center}
\end{table}

\section{Conclusions}

We have performed a calculation of the branching fractions for the semileptonic
decays of $B$ meson into final states containing the narrow orbitally excited
charmed mesons.

We worked in the framework of the constituent quark model of
Ref.~\cite{Vijande2005}. We have calculated the semileptonic decay rates within
the helicity formalism of Ref.~\cite{ivanov} and following the work in
Ref.~\cite{hnvv06}. The strong decay widths have been calculated using two
models, the $^{3}P_{0}$ model and a microscopic model based on the
quark-antiquark interactions present in the CQM model of
Ref.~\cite{Vijande2005}.

From the experimental point of view, Belle and BaBar Collaborations provide
their most recent measurements for the $B$ meson in Refs.~\cite{belle}
and~\cite{aubert09}, respectively.

Our results for $B$ semileptonic decays into $D_{1}(2420)$ and $D_{2}(2460)$ are
in good agreement with the latest experimental measurements by the BaBar
Collaboration.

%%%%%%%%%%%%%%%%%%%%%%%%%%%%%%%%%%%%%%%%%%%%%%%%%%%%%%%%%%%%%%%%%%%%%%%%%%%%%%%%
% acknowledgements (optional)
\acknowledgements{%
This work has been partially funded by the Spanish Ministerio de Ciencia y
Tecnolog\'ia under Contracts Nos. FIS2006-03438,  FIS2009-07238 and
FPA2010-21750-C02-02, by the Spanish Ingenio-Consolider 2010 Programs CPAN
CSD2007-00042 and MultiDark CSD2009-0064, and  by the European
Community-Research Infrastructure Integrating Activity 'Study of Strongly
Interacting Matter' (HadronPhysics2 Grant No. 227431). C. A. thanks a Juan de la
Cierva contract from the Spanish  Ministerio de Educaci\'on y Ciencia.
}

%%%%%%%%%%%%%%%%%%%%%%%%%%%%%%%%%%%%%%%%%%%%%%%%%%%%%%%%%%%%%%%%%%%%%%%%%%%%%%%%
% bibliographic items can be constructed using the LaTeX format in SPIRES
% see http://www.slac.stanford.edu/spires/hep/latex.html
% SPIRES will also supply the CITATION line information; please include it

%
%%%%%%%%%%%%%%%%%%%%%%%%%%%%%%%%%%%%%%%%%%%%%%%%%%%%%%%%%%%%%%%%%%%%%%%%%%%%%%%%

}  % do not remove

%%% Local Variables: 
%%% mode: latex
%%% TeX-master: "../hadron2011.tex"
%%% End: 

\end{document}